\documentclass[manuscript]{aastex}

\begin{document}

\title{Polars Changing State: Multiwavelength Long Term Photometry and Spectroscopy of QS Tel, V834 Cen, and BL Hyi}

\author{Jill R. Gerke\altaffilmark{1}}
\author{Steve B. Howell\altaffilmark{2}}
\author{Frederick M. Walter\altaffilmark {3}}

\altaffiltext{1}{Department of Astronomy, University of Arizona,
  Steward Observatory, 933 N. Cherry Ave., Tucson, AZ 85721;
  jgerke@email.arizona.edu} 
\altaffiltext{2}{WIYN Observatory \& NOAO, P.O. Box 26732, 950
  N. Cherry Ave., Tucson, AZ 85726-6732; howell@noao.edu}
\altaffiltext{3}{Department of Physics and Astronomy, Stony Brook
  University, Stony Brook, NY 11794-3800}

\begin{abstract}
Long term optical and near-infrared photometric and blue spectroscopic
observations were obtained for QS Tel, V834 Cen,
and BL Hyi.  The optical light
curves of all three polars displayed large magnitude changes during our
observations. 
These same high/low state transitions were also apparent
in near-infrared $JHK$ photometry, though with decreased amplitude.
The color of the polar with
respect to its state was examined and found not to be a
good indicator of the instantaneous state.  During low to high state
transitions, a
nearly constant magnitude difference was observed in all three polars. 
This $\Delta$m value was found to be consistent with the level expected
to occur if accretion onto the white dwarf reached the Eddington 
luminosity during the high
state. The high state Balmer decrement was measured for each star and 
used to estimate that the temperature of the emission line forming
region was $\sim$12,000K with N$_H$ near 12.8 dex.  No relationship
between the Balmer emission line strength and the white dwarf magnetic
field strength was seen, in contrast to a good correlation between
these two parameters observed for UV emission lines. \end{abstract}


\newpage

\section{Introduction}
Polars, or AM Her stars, are a subclass of cataclysmic variable
consisting of a close, interacting binary system containing a white
dwarf primary having a magnetic field on the order of 10-200 MG
(Ramsay et al. 2004).  The white dwarf is paired with a low-mass main
sequence secondary and both stars are tidally locked with the binary period
which is generally between 80 minutes and 4 hours (Warner 1995). The
secondary acts as a mass
donor for the white dwarf, overflowing its Roche lobe and transferring
mass at the inner Lagrangian point.

The transfered mass, unable to form an accretion
disk due to the white dwarf's high magnetic field (Araujo-Betancor et
al. 2005), instead forms an accretion column.  The  infalling matter,
funneled by the magnetic field, forms a shock region above the
magnetic pole. The material then continues, at a lower velocity, to
the white dwarf surface where it accretes over a small area covering only a few
thousand square kilometers near the magnetic pole (Howell et
al. 1999).  The shock region and the post shock flow to
the white dwarf surface comprise the accretion column (Cropper 1990). 

Due to the lack of an accretion disk around the white dwarf, the rate
of accretion onto the primary is equal to the instantaneous
rate of mass loss ($\dot M$) by the secondary star (King \& Cannizzo 1998).
Consequentially, polars vary in luminosity in direct relation to
$\dot M$. This is different from other
cataclysmic variables that have accretion disks, wherein the effects of
a changing rate of mass transfer from the secondary star are more subtle.

The high/low state luminosity variation of a polar results
from differing
rates of accretion onto the primary and can be traced
back to differing rates of mass transfer from the
secondary. We note that polars can also show intermediate states (Howell et
al. 2002).  The duration of a state can be a matter of weeks,
months or years.  Although it is uncertain whether conditions on the
secondary actually cause the low state, star spots 
or stellar activity in general, could be responsible
for the decrease in accretion (Araujo-Betancor et al. 2005).
 During a high state, the mass
transfer rate in polars is typically near 10$^{-10}$M$_{\odot}$ 
yr$^{-1}$ (Howell et
al. 1999). During a low state, even though mass transfer due to
Roche lobe overfill may end completely (Howell et al. 1999),
accretion via winds from the
secondary is possible (Kafka et al. 2005).

During the high state, the emission from the X-ray through the optical and IR
bands is dominated by accretion flux.  The
difference in optical magnitude between a high state and low state of a
polar is typically near 2-2.5 magnitudes. For example, Kafka et al. (2005)
showed that AM Her changed approximately 2 mag between states and
out of fourteen polars surveyed by Ramsay
et al. (2004), twelve had a change in magnitude within the
1.5-3.0 mag range; the average change was 2.1 $\pm$.2 mag.  Despite any
differences in the white dwarf's magnetic field strength or other
parameters governing the polar, the comparable change in luminosity at
the change of state suggests there is some similarity in the cause and
behavior of the high state luminosity emitted at or near the accretion
region of polars.  

In order to understand this similarity in luminosity change and
the behavior of the change of state of polars we undertook a long
term study involving observations from the SMARTS\footnote{
SMARTS, the Small and Medium Aperture Research Telescope Facility, is
a consortium of universities and research institutions that operate the
small telescopes at Cerro Tololo under contract with AURA.}
facilities. Our observations consisted of optical (B, V and I) and
infrared (J, H and K) light curves along with blue (3500-5300\AA)
spectra of several polars.  

In the following sections, examination of the optical light curves and
blue spectroscopic data obtained from the
study for QS Tel, BL Hyi and V834 Cen leads us to propose a model
where the accretion region on the white dwarf 
reaches the Eddington luminosity in the high state.  
We note that system color is not dominated by high/low state conditions,
in the optical or the near-IR.
From the spectroscopic data, the Balmer decrement
(H$\gamma$/H$\beta$) is determined for the high state of a polar
and used to find the high state temperature and
density of the emitting material. We also examine 
the emission line strengths with respect to the strength of the white dwarf
magnetic field.

\section{Observations and Data Reduction}
All the data we report herein were obtained with the SMARTS
facilities at Cerro Tololo by the SMARTS service observers.
All dates below refer to the civil date at the start of the night.  In
addition to QS Tel, BL Hyi and V834 Cen, we also monitored EF Eri and
VV Pup.  These two polars, however, remained in the low state for the
duration of this study and therefore are not discussed further here. 

\subsection{Imaging}
We obtained images of the targets using the
ANDICAM\footnote{
  http://www.astronomy.ohio-state.edu/ANDICAM/detectors.html }
dual-channel photometer on the 1.3m telescope.
The optical detector is a 1024x1024 CCD; the near-IR channel uses
a Rockwell 1024x1024 HgCdTe Hawaii array. The optical and near-IR
channels operate independently, so we can take a set of near-IR images
while integrating in the optical.
 
We obtained data on many nights between 2003 August and 2005 January (see
Table 1). The optical data consisted of single images through B, V,
and I filters. Integration times were 100 seconds per filter
(200 seconds for QS Tel). The near-IR images consisted of three dithered
27~second integrations through each of the J, H, and K (CIT/CTIO) filters.
Images are obtained sequentially through each of the filters, with the
$B$ and $J$, $V$ and $H$, and $I$ and $K$ pairs obtained simultaneously.
QS Tel was too faint to obtain reliable $JHK$ photometry.

The optical ANDICAM data are processed (overscan subtraction and
flat-fielding) prior to distribution.
Because we obtained only single images through each filter each night,
the noise is dominated, in some cases, by cosmic rays and hot pixels.
Aside from a 2x2 rebinning, the IR data are delivered raw.
We generate the flat field images from the
dome flats obtained about every other night.
We generate a sky image by taking the median of the dithered images.
We subtract the sky image from the individual frames, then shift and add the
individual frames.
 
Standard practice is to use ANDICAM for differential photometry,
but the SMARTS project routinely obtains images of optical and near-IR
standard fields on photometric nights to allow defined zero-points for
the photometry.
Our optical light curves consist of relative 
magnitudes and the color measurements are instrumental. 
The near-infrared
values, however, are calibrated magnitudes, so we quote true color values. 
 
\subsection{Spectroscopy}
The spectra (Table 1) were obtained with the RC spectrograph
on the 1.5m telescope. This is a slit spectrograph, with a 300\arcsec\
long slit oriented E-W. The stars are always observed through a 110$\mu$m
(1.5\arcsec) slit and obtained three images at each epoch in order to
filter cosmic rays.  Each set of images is
accompanied by a wavelength calibration arc lamp exposure.
We have developed a pipeline, written in IDL, to process the data.
The pipeline subtracts the overscan, and divides by the normalized flat
field image.

A median image is generated from the three images and we extract the spectra,
both by using an unweighted boxcar extraction and by fitting a Gaussian profile
at each wavelength. For the boxcar extraction, we determine the width of the
extraction box by fitting a Gaussian to the spatial profile, and the
background is measured to either side of the source. In the Gaussian extraction
we fit a Gaussian to the spatial profile at each wavelength, and retain the
number of counts in the fit. This is less sensitive to noise in individual
pixels, and provides a higher S/N spectrum.

A spectrophotometric standard, either LTT 4364 (Hamuy et al. 1992,
1994) or Feige 110 (Oke 1990; Hamuy et al. 1992, 1994),
was observed each night in order to convert the counts spectrum to a
flux spectrum.  We note that there will be systematic deviations at the
blue end of the spectrum when the targets and standard are observed at
different air masses, because the slit orientation is fixed and not free to
rotate with the parallactic angle. This effect is largest for BL Hyi, but
has no significant effect on any of our conclusions.
Due to seeing-related slit losses, we do not obtain
absolute fluxes, but rather use the standard star to recover the shape
of the continuum and provide relative fluxes.

Most of the spectra were obtained with grating 26 in first order, which
provides 4.3\AA\ resolution between 3600 and 5200\AA.
Some spectra were taken in first order using grating 47 and the GG~495 order
sorting filter to obtain 3.1\AA\ resolution between 5650 and 6970\AA.
This resolution is sufficient to resolve the H$\alpha$ line.

\section{Photometric Analysis}
 Figure 1 shows the optical light curves for QS Tel, BL Hyi and V834
 Cen, with both the state
of the system and the boundaries used to define those
states indicated.  The luminosity boundaries seen in Figure 1 for the states of
the polars were not difficult to determine, as each
polar had a baseline low state luminosity with only small
variations due to the heterogeneous nature of the accretion stream or
flares from the secondary (see Ramsay et al. 2004; Howell et al. 2002; Kafka
et al. 2005 for further explanation).  To compensate for flares and
for observations taken when the polar was changing states, we placed
some of the photometric measurements in a ``transitional region'' between
the well defined high and low states for QS Tel and V834 Cen (see
Figure 1).  The change of state for BL Hyi occurs between observations
 and does not show transitional points. However, the high state does contain a
 variation of about 0.2 magnitudes caused by an
 orbital modulation as the active pole rotates in and out of view (Wolff 1999).

Figure 2 and Figure 3 show the J, H and K band light curves for BL Hyi and V834
Cen, respectively. The near-infrared light curves show 
brightness changes coincident with, though of lesser amplitude than, the high/low optical light curve changes.
This correlation 
indicates that accretion flux can be a major 
factor in the near-infrared brightness of a
polar, despite the presence of a red secondary. 
Orbital modulation is visible in the near-IR high state light curves 
of BL Hyi just as it was in the optical.  

As Figure 1
shows the change in B magnitude between the low state and the high
state for each polar is $\sim$2-2.5 magnitudes.  As stated above, this 
level of brightness change is similar
to that observed for many
other polars as well.  In order to determine a possible explanation for the
consistency of the change in luminosity between high and low states in
polars, the
Eddington luminosity was considered as a possible limiting
factor for the luminosity of the accretion process (see \S3.2).

\subsection{Color of the Polars}

The SMARTS photometric observations were used
to create color-color plots where the state of the
polar at the time of the observation is indicated (Figures 4 \& 5).
We noted the state of each observation in order to determine whether
color could be used as a good indicator of the accretion state. 
While there is some
connection between color and state, the results are not consistent for
all polars. 
 
The optical data of QS Tel is grouped according to state in the
color-color plot.  The SMARTS observations show QS Tel
becoming progressively bluer as it moves from the low to the
transitional and into the high state.  These
color groupings confirm the accretion
region dominates the optical light in the high state causing the system to
appear more blue.  In the low state, some
spectral combination of the white dwarf and the secondary star occurs
resulting in a redder color.  During this time, the red secondary
contributes a higher fraction of the optical light while the cooling accretion
region contributes less.
The relatively small range in 
color for QS Tel may be the result of the hot white dwarf
(T$_{eff}$=17500K, Rosen et al. 2001) dominating the optical in
the low state as well as the high state.

The spread of the data according to high and low state for V834 and BL
Hyi (Figures 4 \& 5) suggest a more complex spectral relationship
between the white dwarf and the secondary as the system goes from a
high state to a low state. 
In the optical, the transitional state of V834 Cen is redder than
either the low or high state.  While the high state appears to be
slightly bluer than the low state, there is some overlap.  The
overlap of the range of colors for the high and low states could be a result
of the secondary and the white dwarf both contributing to create a
complex spectral energy distribution in the low state.  In the
near-IR, however, there is less overlap in the color between states
and the color of the transitional state separates the bluer color of
the high and redder color of the low state.    

The results from BL Hyi also appear to be complex. 
The high state has a range of colors from red to blue while the
low state observations all show a similar color that is
bluer than the color of the high state in (V-I)$_i$, but redder
in (B-V)$_i$. The high
state data of BL Hyi has a larger spread toward the red than is seen
with the other systems, likely caused by color modulation over
the orbital period due to self eclipse of the accretion region, where
the amplitude has been shown to increase with
increasing wavelength (Warner 1995).  This argument is supported by the near-IR
color plot which also shows the high state covering a range from red
to blue.

In the above discussion of high and low state color, we have not yet
added effects due to cyclotron humps. During low states, these
wide ($\sim$1000\AA) continuum humps can provide strong effects at
specific wavelengths. The location and strength of cyclotron humps
vary from system to system as well as over orbital phase.  The
amplitude of the humps depends mainly on the mass transfer rate, with
the largest amplitudes generally occurring during the low state.
Given the range of the observed colors in the high and low states, we
conclude that the color of a polar can not be used alone as a good
indicator of its accretion state.     

\subsection{Eddington Luminosity}

The Eddington luminosity for accretion onto the white dwarf in each of our
observed polars has been calculated using stellar parameters
taken from the literature (see Table 2).  The
Eddington Luminosity (L$_E$) is given by 
$$L_E=\frac{4\pi GcM_{WD}}{k}$$  
where G is the gravitational constant, M$_{WD}$ is the mass of the white
dwarf and c is the speed of light.  The opacity of the
accreting matter (k) was estimated to be 1.2 cm$^{2}$g$^{-1}$, assuming a
hydrogen mass fraction of 1 (Warner 1995). 

 The change in state of a polar is due to changes in the mass transfer
 rate, however material only accretes over a small region near the
 magnetic pole, the accretion region. Thus, only that fractional area
 of the white dwarf, estimated to be $\sim$10\% of the surface area of
 the white dwarf, changes luminosity appreciably.  Due to the
 similarity in white dwarf mass and composition, the Eddington
 luminosities will also be similar.  This similarity results in a
 change in magnitude at the change of state that is approximately
 uniform for many polars. 

While the accretion region
reaches its peak luminosity in the high state due to the heightened
rate of mass transfer, the luminosity of the
rest of the white dwarf remains essentially unchanged from that of the
low state.  The
resulting high state luminosity (L$_H$) of the polar can be estimated
as L$_H$=.9L$_P$ + .1L$_{PE}$ + L$_S$ where L$_P$ is the low state
luminosity of the primary, L$_{PE}$ is the estimated Eddington
luminosity of the primary, and L$_S$ is the luminosity of the
secondary. The coefficients refer to the percentage
of the surface area of the star with that luminosity.  In
the low state, the accretion region
receives a negligible amount of matter,
therefore the accretion region is not heated measurably and the entire
surface area of the primary has approximately the same luminosity
L$_L$=L$_P$ + L$_S$ where L$_L$ is
the low state luminosity. To summarize, our toy model proposes that
during the low state the entire white dwarf has nearly the same luminosity
(i.e., a uniform surface temperature),
while during the high state the luminosity of the accretion region
increases to the Eddington luminosity and the luminosity of the rest
of the star remains unchanged. 

Using the above expressions for the high state and low state
luminosities of a polar, the change in luminosity due to a change to a high
state is determined. The low state luminosity of the primary is
estimated by the low state luminosity of the system, L$_L$=L$_P$
(assuming an optically dim secondary) and the change in
luminosity is estimated as .1L$_{PE}$-.1L$_L$.  The
Eddington luminosity was then converted to an apparent magnitude using
the polar distances from Table 2 and the predicted change in
magnitude was calculated with the results shown in the final column of Table 2.

The light curves of BL Hyi, QS Tel and V834 Cen (Figure 1) show a
change of about 2 magnitudes, however, our simple model predicts a slightly
smaller change in luminosity for these polars.  The Eddington
luminosity gives the
maximum radiative luminosity that can be supported by an object with a
stable, isotropic, homogeneous and fully ionized atmosphere (Moon et
al. 2003).  These atmospheric conditions do not strictly apply to the white
dwarf of a polar since the accretion column is inhomogeneous
and covers a small portion of the white dwarf.  In addition, the 
accretion region does
not cool off instantaneously to the white dwarf temperature. We 
have also ignored other details such as 
possible emission through the sides of the column or from the accretion stream.

The realistic, complex details of
the atmosphere above the accretion region may make our calculations an
underestimate of the change in magnitude due to the change to a high
state. Also, the size of the accretion region could be
overestimated (Warner 1995), which would in turn cause the change in
magnitude to be overestimated, plus some polars may have more than one
active accretion region at one time.  Finally, by assuming an
optically dim secondary, the change in magnitude is
underestimated by one tenth the optical magnitude of the secondary star
(0.1L$_S$).

\section{Spectroscopic Analysis}
A representative SMARTS spectrum from the high state and the low state of
BL Hyi can be seen in Figures 6 \& 7, respectively.  The high
state and low state are easily distinguishable as the high state
has discernible, strong emission lines, while the low state does not. 

\subsection{H$\gamma$/H$\beta$ Decrement}
By calculating the Balmer decrement (H$\gamma$/H$\beta$) for the high
state spectra,
ranges for the
temperature and density within the emission line forming region can be
estimated.  Williams~(1991) modeled the
Balmer decrement for emission lines in cataclysmic variables using
grids of specific temperature, density, and inclination. While
Williams's work was for line forming regions in accretion disk chromospheres, the same
temperature-density relations should hold for the line forming regions
in the accretion stream, a chromosphere-like area above and near an
accretion pole.  William's Balmer decrements, averaged over
inclination, are shown in Figure 8.  Figure 9 plots the high state decrements
determined from our SMARTS polar spectra over a time interval of about
500 days. 

Using the optical data (Fig. 1) to define the state, we calculated the
average Balmer decrement for the high state of each polar.  Table 3
gives our results for the H$\gamma$/H$\beta$ decrement for each star, along with the estimated temperatures and number
density of hydrogen drawn from Williams~(1991). The high state emission line
forming regions of the polars are found to be similarly hot
(10,000-15,000K) and dense (12.4-13.2 dex). This is in good agreement with Warner (1995).

\subsection{Emission Line and Magnetic Field Strength}

Using high state spectroscopic observations of polars obtained by the 
{\it International Ultraviolet Explorer} (IUE), 
Howell et al.~(1999)
compared the strength of several UV emission lines to the
magnetic field strength of the white dwarf for a number of polars.  They
found that the UV emission line strength decreased with increasing
magnetic field strength. The same process used by Howell et al.~(1999) for UV
emission lines was applied to our current spectroscopic dataset.

Using spectra obtained on nights listed as clear (to avoid decreased
emission line strength due to clouds) an average high state emission line flux
was determined for H$\beta$, H$\gamma$, and H$\delta$ for each of our
three polars. The average emission line fluxes were then converted
into average line luminosities, using the distances given in Table 2, with our
results compiled in Table 4.  Comparing line
strength to the magnetic field strength of each polar, we find that
the trend of decreasing line strength with increasing magnetic field
seen in the UV is not evident in the Balmer lines.  

Within the uncertainties, all three polars show
similar strength Balmer emission lines.  
This inconsistency with the results of Howell et al.~(1999) likely 
indicates that the strength of the
optical emission lines are less directly affected by the
magnetic field strength compared with those in the UV.
The Balmer emission is likely to be more effected by the local
conditions under which they form in the accretion stream and column.
The location of the emission line forming regions must also be
considered.  The UV lines form lower in the
accretion column, nearer to the magnetic pole. 
The optical emission lines, on the other hand, form in a
region in the accretion stream with a weaker magnetic
field and an environment of varying temperature and density.

\section{Conclusion}

The lack of an accretion disk around the white dwarf in a polar directly
relates the rate of mass transfer from the secondary to the rate of mass
accretion onto the primary.  Therefore, when mass transfer from the
secondary slows, accretion onto the primary also slows and the
luminosity of the polar changes.  A change of state
is generally characterized by a change in the optical brightness of $\sim$2-2.5
magnitudes.  The consistency in magnitude change at the change of
state for most polars is consistent with 
the accretion region reaching the Eddington luminosity
in the high state. 

The color-color relations in both the optical and near-IR show different
color trends for each polar over the different states.  The optical
color of QS Tel shows a polar which is blue in the high state, redder in the
low state, with the transitional region falling between the two.
Colors in both the
 optical and near-IR for V834 Cen and BL Hyi show a
complex relationship between high and low states. 
The near-IR light curves also show a change in
luminosity at the change of state, however the luminosity change is
smaller than in the optical and decreases when moving from the J to
the K bands. 

The Balmer decrement (H$\gamma$/H$\beta$) determined from high state
spectroscopy, reveals that the emission line forming regions are hot
(8000-15000K) and fairly dense (Log N$_H$=12.4-13.2).  We also
measured the Balmer line emission strength, finding no clear
relationship to the magnetic field strength of the polar.  In order to
compare the states of a polar and to determine how parameters change
during a state change, long term studies are needed.  These systems
are dynamic and must be monitored to increase understanding of their
properties.  

\acknowledgments
We wish to thank the SMART consortium for their kind allocation of telescope
time and the SMART observers for their diligence and hard work in obtaining
the data used herein. We would also like to thank the anonymous
referee for their helpful comments and suggestions. 

\newpage

\begin{deluxetable}{rrl|rrl}
\tablecolumns{6}
\tablewidth{0pc}
\tablenum{1}
\tablecaption{Observation Log}
\tablehead {
\colhead{} & \multicolumn{2}{c}{ANDICAM} & \colhead{} &\multicolumn{2}{c}{Spectrograph}\\ 
\cline{2-3} \cline{5-6}
\colhead{Star} & \colhead{Nights} & \multicolumn{1}{l}{Span (UT)} & \colhead{} & \colhead{Nights} & \multicolumn{1}{l}{Span (UT)}}
\startdata
QS Tel   & 73 & 2003 Aug 21  - 2004 Oct 28 & & 38 & 2003 Aug 14 - 2005 Jun 03 \\
BL Hyi   & 74 & 2003 Aug 21  - 2005 Jan 26 & & 76 & 2003 Aug 19 - 2005 Jun 03 \\

V834 Cen & 48 & 2004 Jan 13  - 2004 Sep 18 & & 44 & 2004 Jan 20 - 2005 Jun 05 \\
\enddata
\end{deluxetable}

\begin{deluxetable}{lccccc}
\tablecolumns{6}
\tablewidth{0pc}
\tablenum{2}
\tablecaption{Eddington Luminosity Parameters}
\tablehead {
\colhead{Star} & \colhead{Mass of} & \colhead{Distance} & \colhead{Low State} & \colhead{Eddington} & \colhead{$\Delta$ Magnitude}\\
\colhead{ } & \colhead{White Dwarf} & \colhead{(pc) } & \colhead{{\it V} Magnitude} &
\colhead{Luminosity (ergs/s)} & \colhead{ }
}

\startdata
QS Tel    &  0.71 M$_{\odot}$\tablenotemark{1}   &  170\tablenotemark{1}   &  17.3\tablenotemark{4}   &  6E37   &  1.7\\
Bl Hyi    &  0.50 M$_{\odot}$\tablenotemark{2}   &  132\tablenotemark{2}   &  17.4\tablenotemark{5}   &  4E37   &  1.7\\
V834 Cen  &  0.64 M$_{\odot}$\tablenotemark{3}   &   80\tablenotemark{3}   &  16.9\tablenotemark{5}   &  5E37   &  1.8 \\
\enddata
\tablerefs{(1) Rosen et al. 2001; (2) Wolff et al. 1999; (3) Mauche 2002; (4)Ferrario et al. 1994; (5) Araujo-Betancor et al. 2005.}
\end{deluxetable}
\newpage

\begin{deluxetable}{lccc}
\tablecolumns{4}
\tablewidth{0pc}
\tablenum{3}
\tablecaption{High State Temperature and Density Ranges 
 based on the Balmer Decrement (H$\gamma$/H$\beta$)}
\tablehead {
\colhead{Star} & \colhead{H$\gamma$/H$\beta$} &
\colhead{Temperature (K)} & \colhead{Log N$_H$}
}
\startdata
QS Tel   &  .83$\pm$.04 & 10000  & 13.0 \\
  &   & 15000 & 12.8 \\

BL Hyi   &  .80$\pm$.002 & 10000  & 12.5 \\
  &   & 15000 & 12.4 \\

V834 Cen &  1.01$\pm$.02 & 10000  & 13.2 \\
  &   & 15000 & 13.0 \\
\enddata
\end{deluxetable}

\newpage

\begin{deluxetable}{lcccccc}
\tablecolumns{5}
\tablewidth{0pc}
\tablenum{4}
\tablecaption{Magnetic Field Strength and Optical Emission Line Luminosity}                                      
\tablehead{
\multicolumn{2}{l}{ } &
\multicolumn{3}{c}{Log Average Line Luminosity} \\
\colhead{Star} & \colhead{Magnetic field (MG)}  &  \colhead{H$\beta$} & \colhead{H$\gamma$} & \colhead{H$\delta$}
}
\startdata
QS Tel  &
56\tablenotemark{1} &  29.24$\pm$.09 &  29.29$\pm$.12   &  29.18$\pm$.08  \\

BL Hyi  &
23\tablenotemark{2} &  29.44$\pm$.08  &  29.12$\pm$.19   &  29.07$\pm$.10 \\

V834 Cen &
31\tablenotemark{3} &  29.32$\pm$.16  &  29.30$\pm$.15   &  29.32$\pm$.17 \\

\enddata
\tablerefs{(1) Ferrario et al. 1994; (2) Ferrario et
  al. 1996; (3) Ferrario et al. 1992.}
\end{deluxetable}

\newpage

\begin{figure}
  \includegraphics[angle=270,scale=.75]{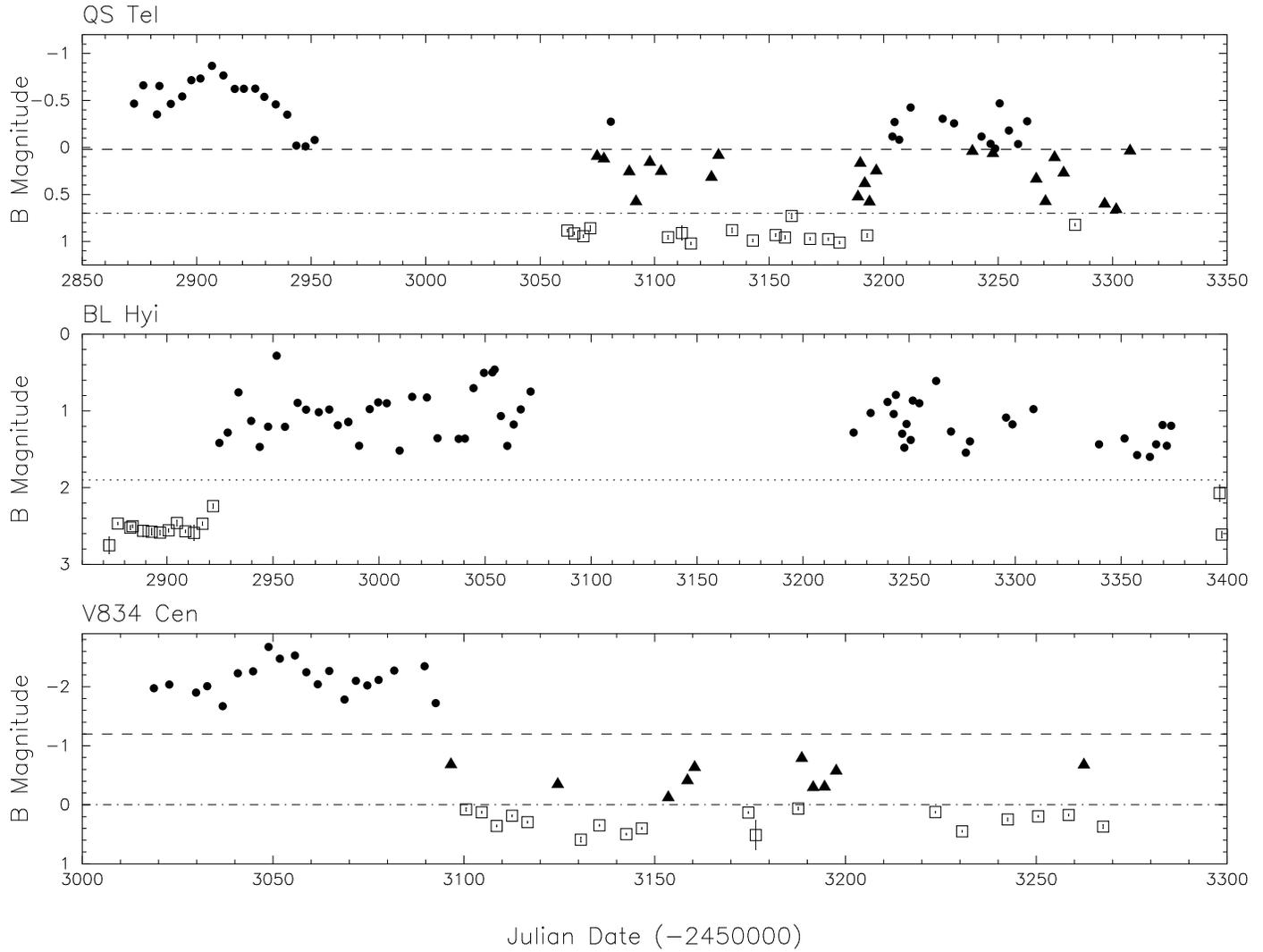}
  \caption{Light curves in relative B Magnitude vs Julian Date.
  ${\blacktriangle}$~Transition;
  ${\square}$~Low state; \large${\bullet}$\normalsize~High state.  The
  dashed line indicates the lower limit to the high states while the
  dot-dashed line indicated the upper limit to the low state.  Points
  between the lines are the transition region.  BL~Hyi shows a clean
  low to high state transition.}
\end{figure}
\newpage

\begin{figure}
  \includegraphics[angle=270,scale=.75]{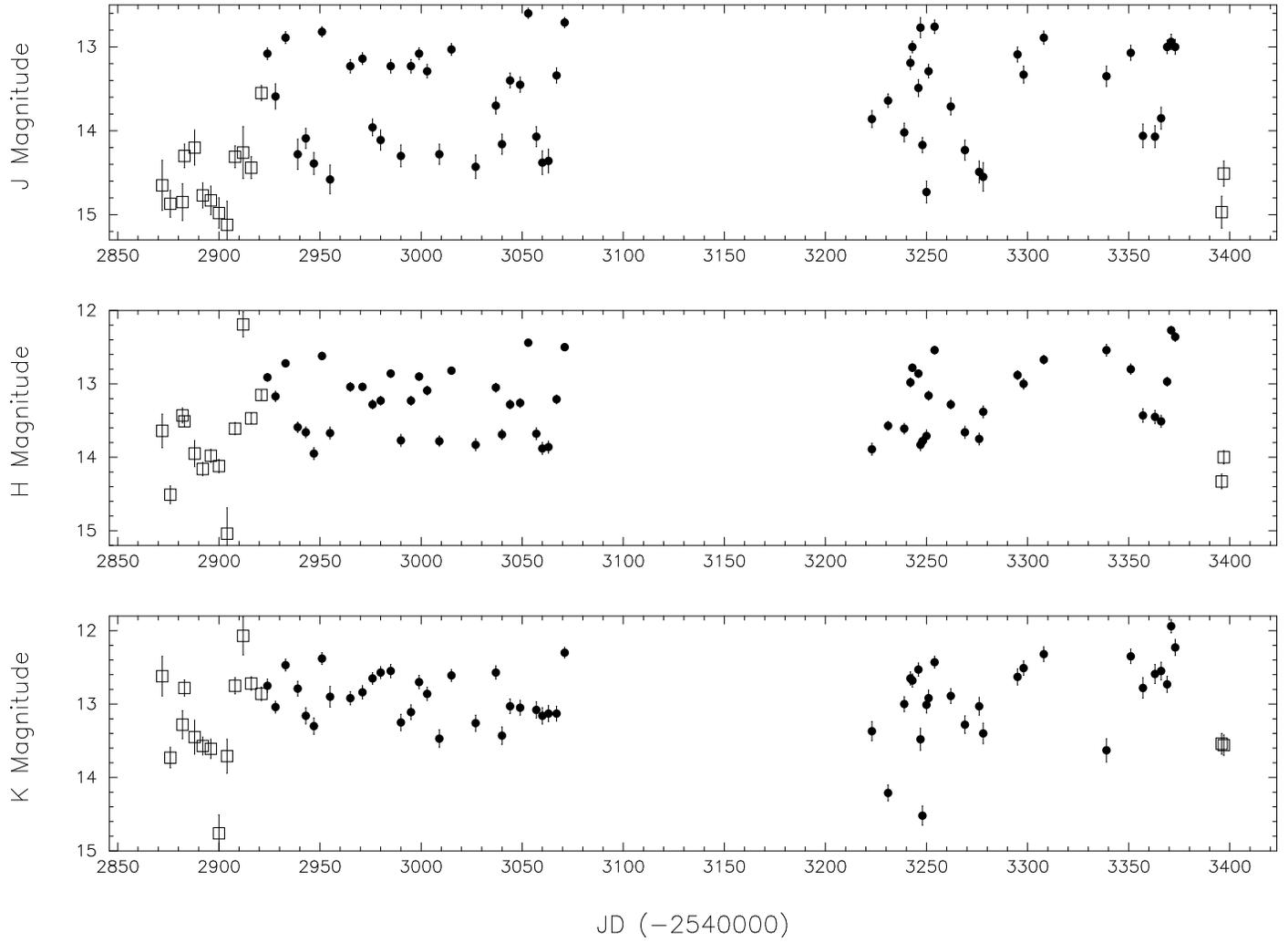}
  \caption{Magnitude of BL Hyi in J, H and K Bands vs Julian Date.  The state of BL Hyi during each observation, as defined by the optical data, is denoted as follows: ${\square}$ ~Low state; \large${\bullet}$\normalsize~High state.}
\end{figure}
\newpage  

\begin{figure}
  \includegraphics[angle=270,scale=.75]{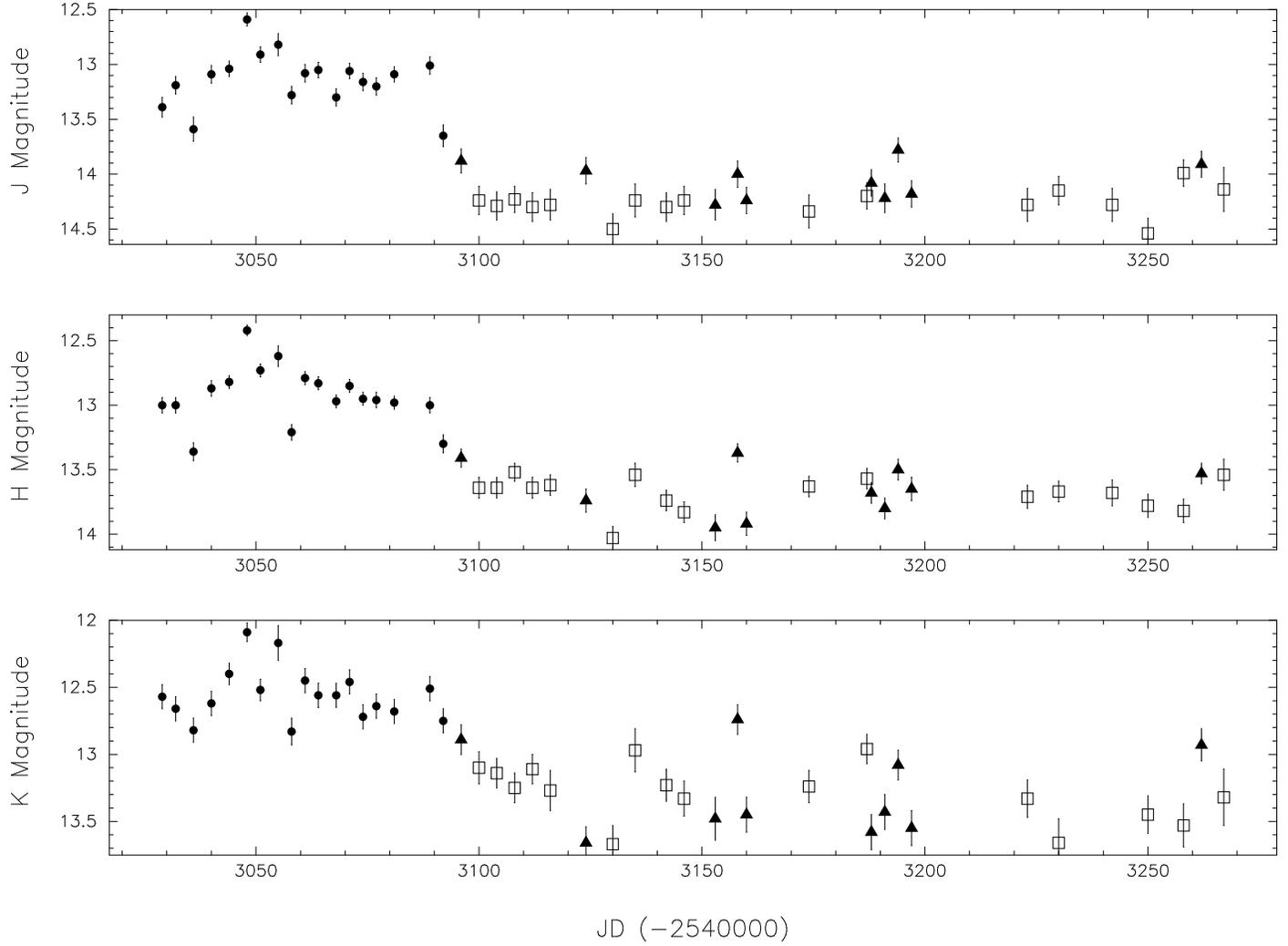}
  \caption{Magnitude of V834 Cen in J, H and K bands vs Julian
  Date. The state of V834 Cen at the time of the observation, as
  defined by the optical data, is denoted as follows:
  ${\blacktriangle}$~Transition; ${\square}$~Low state; \large${\bullet}$\normalsize~High state.}  
\end{figure} 
\newpage

\begin{figure}
  \includegraphics[angle=270,scale=1.0]{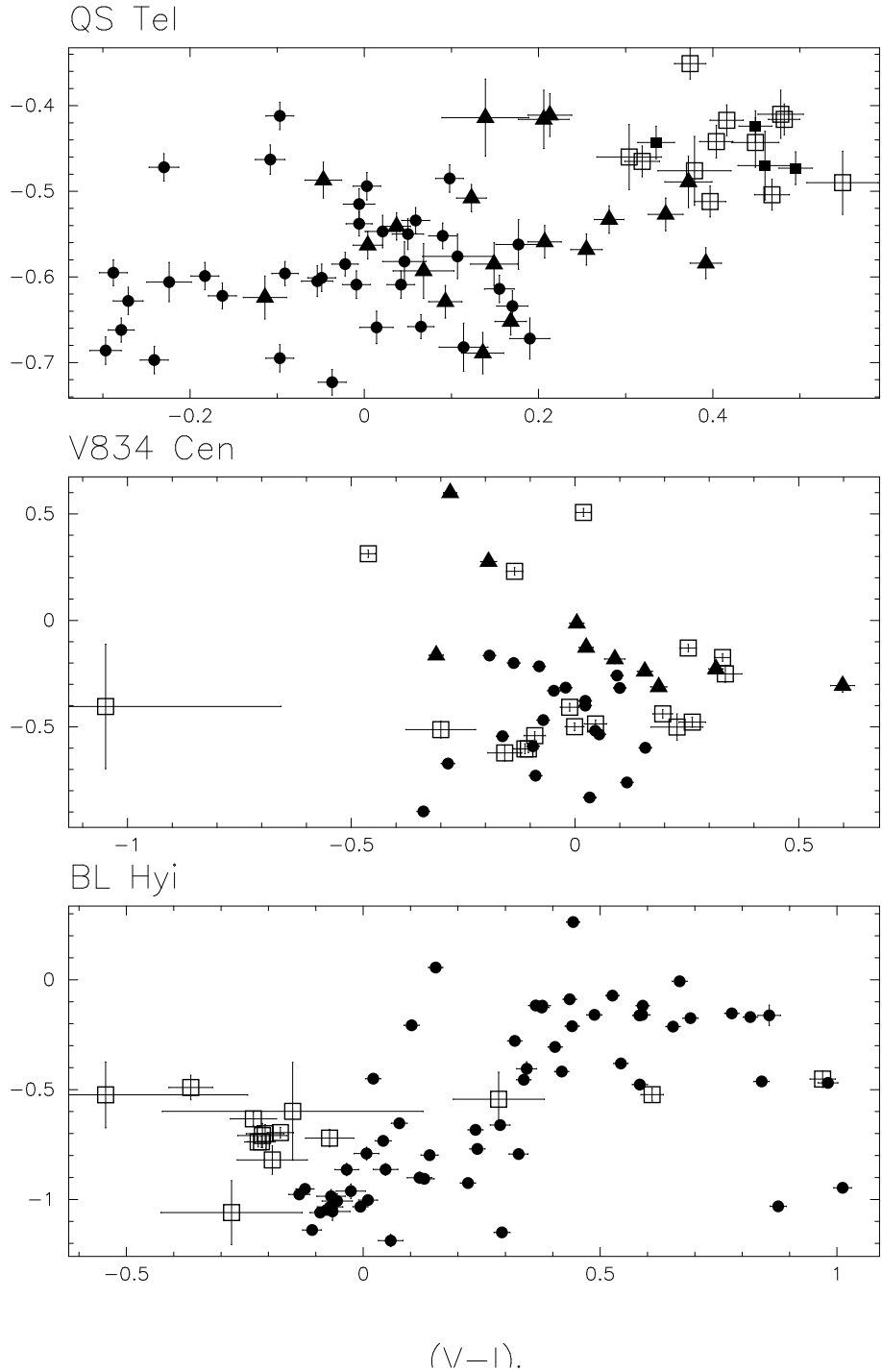}
  \caption{Color-color plots in the optical for the three polars. The
  state of the polar during an observation, as defined by the optical
  data, is denoted as follows: ${\blacktriangle}$~Transition;
  ${\square}$~Low state; \large${\bullet}$\normalsize~High state.} 
\end{figure}

\newpage  

\begin{figure}
  \includegraphics[angle=270,scale=1.0]{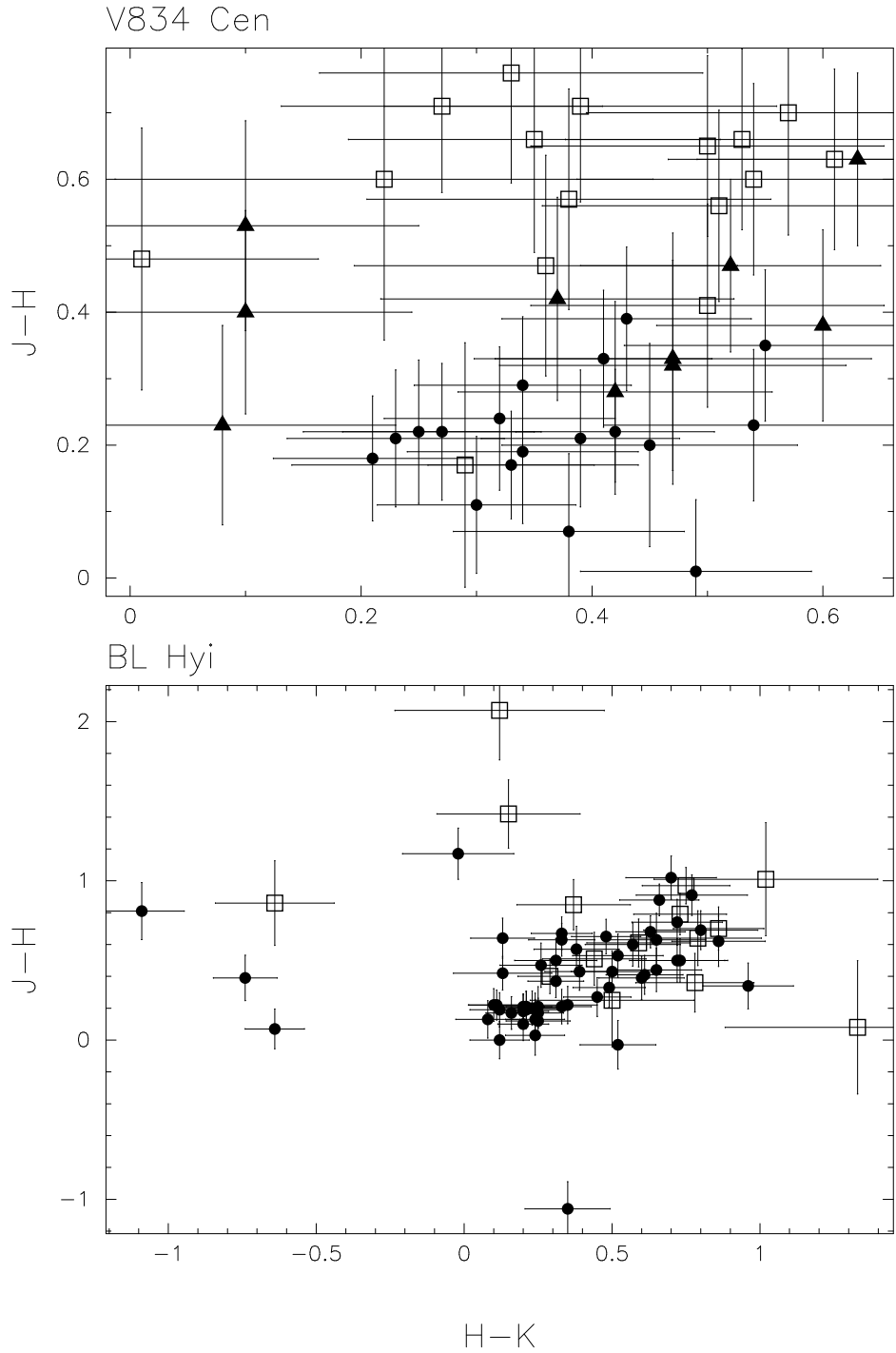}
  \caption{Color-color plots in the near-Infrared for V834 Cen and BL Hyi. The state of the polar during the observation, as defined by
  the optical data, is denoted as follows: 
  ${\blacktriangle}$~Transition; ${\square}$~Low state;
  \large${\bullet}$\normalsize~High state.}  
\end{figure}

\newpage

\begin{figure}
  \includegraphics[angle=270,scale=.55]{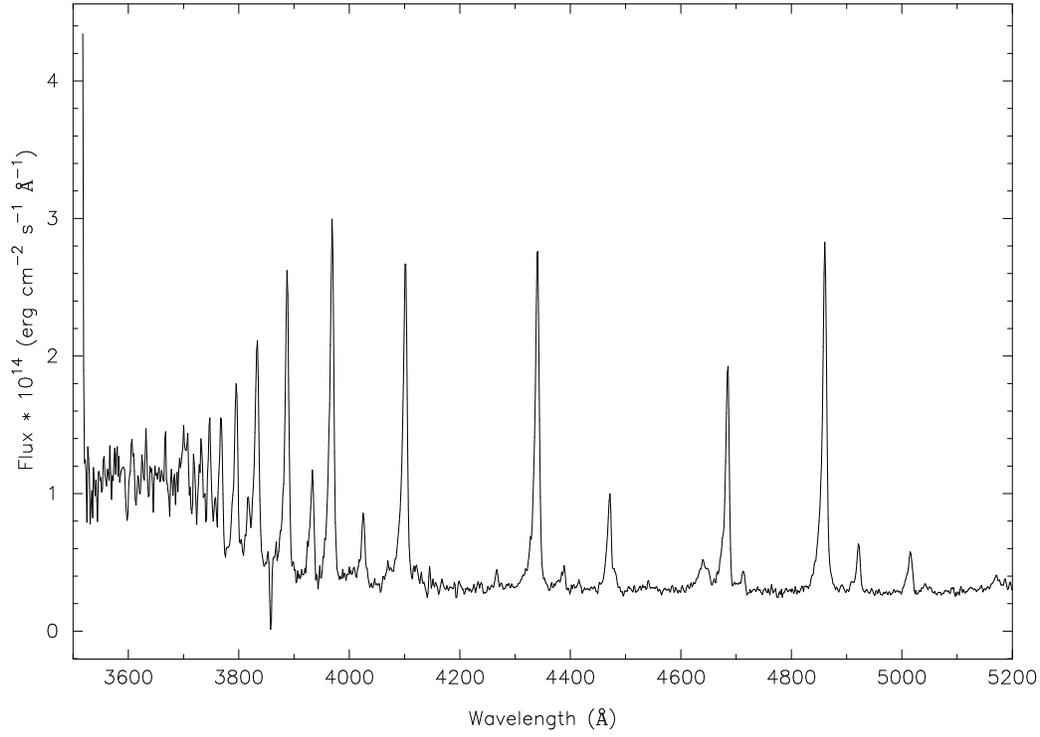}
  \caption{High State spectrum from BL Hyi obtained on 2004 Sept
  15.}
\end{figure} 

\begin{figure}
  \includegraphics[angle=270,scale=.55]{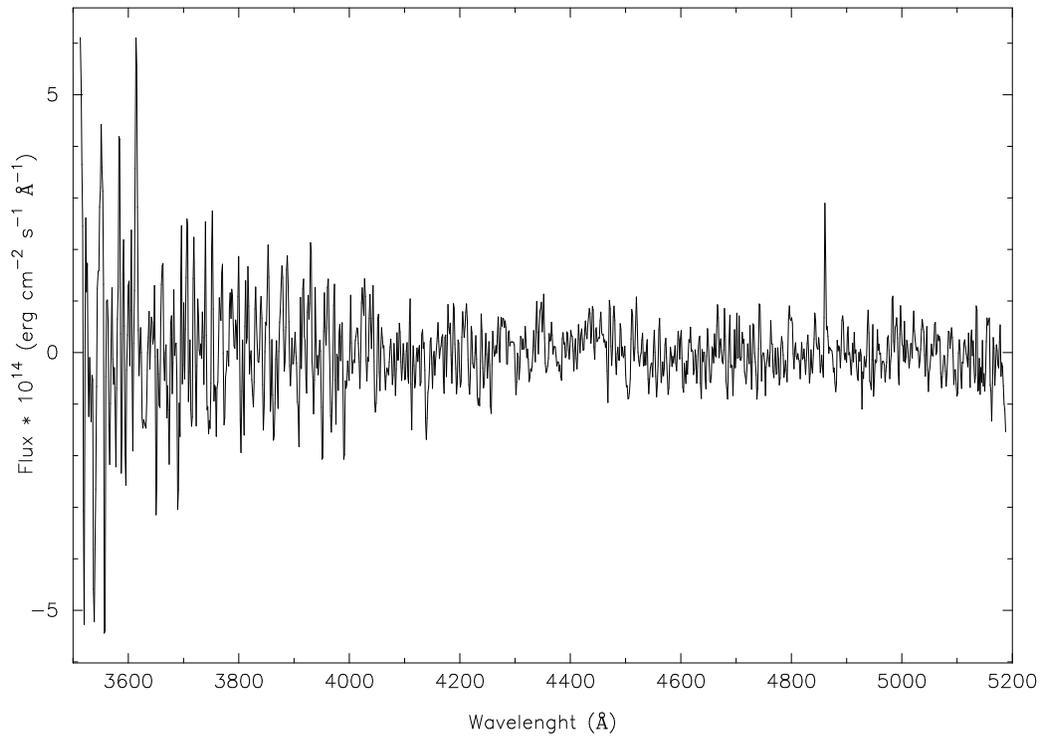}
  \caption{A low state spectrum from BL Hyi averaged from seven observations.} 
\end{figure}

\newpage

\begin{figure}
  \includegraphics[angle=270,scale=.75]{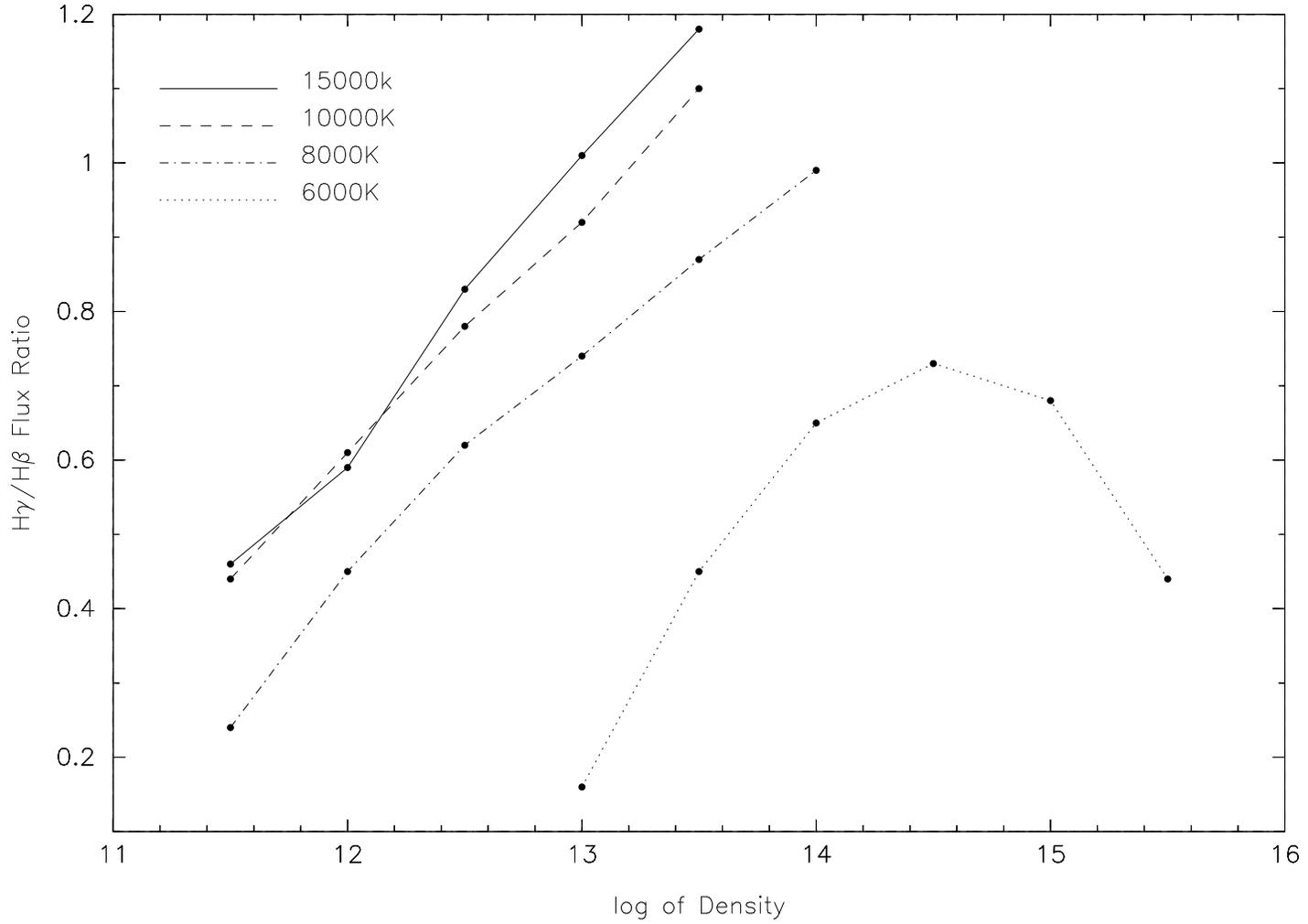}
  \caption{Balmer Decrement (H$\gamma$/H$\beta$) vs Log of Density
  (g/cm$^3$) for various temperatures from  Williams (1991).
  } 
\end{figure}

\newpage

\begin{figure}
  \includegraphics[angle=270,scale=.75]{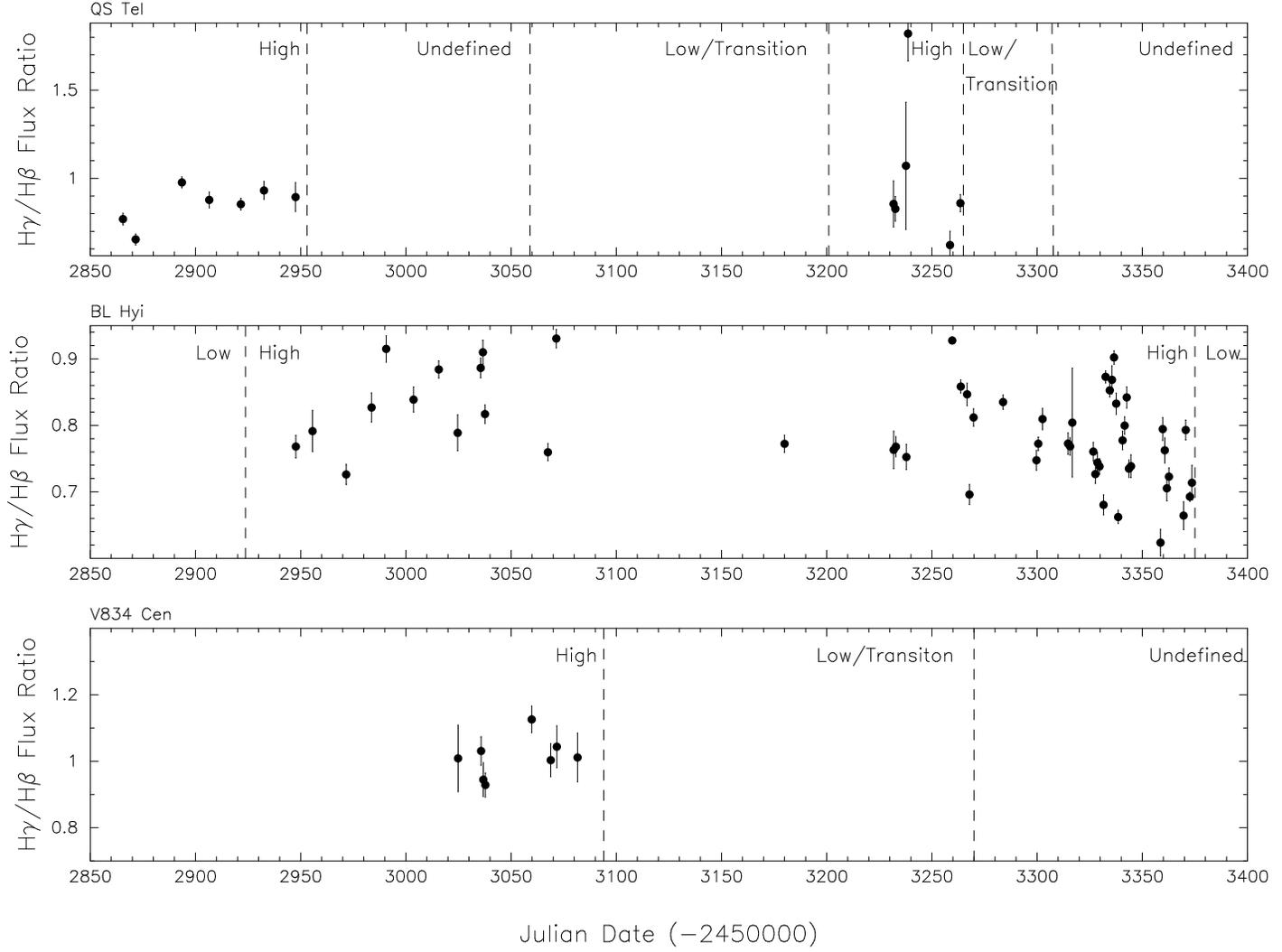}
  \caption{Balmer decrement vs Julian date. The state of the polar is
  marked using the transitions shown in Figure 1. 
  The Balmer decrements are only defined in the high state (see Table 3).}
\end{figure}

\end{document}